\documentclass[journal]{IEEEtran}

\usepackage{cite}
\usepackage{caption}
\usepackage{subfigure}
\usepackage{float}
\usepackage{hyperref}
\usepackage{fontawesome5}
\usepackage{xcolor}
\usepackage{ulem}
\ifCLASSINFOpdf
  \usepackage[pdftex]{graphicx}
\else
\fi
\usepackage{amsmath}
\usepackage{authblk}
\usepackage{marvosym}

\usepackage{stfloats}

\newcommand{\orcid}[1]{\href{https://orcid.org/#1}{{\faOrcid}}}

\begin{document}

\title{Explainable AI for Autism Diagnosis: Identifying Critical Brain Regions Using fMRI Data}

\author{Suryansh~Vidya \textsuperscript{1},
        Kush~Gupta \textsuperscript{1},
        Amir~Aly \textsuperscript{1}\textsuperscript{\faEnvelope[regular]},
        Andy~Wills \textsuperscript{1},
        Emmanuel~Ifeachor \textsuperscript{1}
        and~Rohit~Shankar \textsuperscript{1,2}% <-this % stops a space
%\thanks{Manuscript received XX,XXXX; revised XX,XXXX}

% \thanks{\textsuperscript{1}Suryansh Vidya, Kush Gupta, Amir Aly, Andy Wills, Emmanuel Ifeachor, and Rohit Shankar are with the University of Plymouth. (e-mail: \href{mailto:suryansh.vidya@students.plymouth.ac.uk}{SV}, \href{mailto:kush.gupta@plymouth.ac.uk}{KG}, \href{mailto:andy.wills@plymouth.ac.uk}{AW}, \href{mailto:E.Ifeachor@plymouth.ac.uk}{EI}, \href{mailto:rohit.shankar@plymouth.ac.uk}{RS}).}

\thanks{\textsuperscript{1}Suryansh Vidya, Kush Gupta, Amir Aly, Andy Wills, Emmanuel Ifeachor, and Rohit Shankar are with the University of Plymouth. (\texttt{e-mail: suryansh.vidya@students.plymouth.ac.uk; kush.gupta@plymouth.ac.uk; amir.aly@plymouth.ac.uk; andy.wills@plymouth.ac.uk; E.Ifeachor@plymouth.ac.uk; rohit.shankar@plymouth.ac.uk}).}

\thanks{\textsuperscript{2}Affiliated with Cornwall Partnership NHS Foundation Trust, Truro, UK}

% \thanks{\textsuperscript{*}Corresponding Author: Amir Aly, University of Plymouth. Email:  \href{mailto:amir.aly@plymouth.ac.uk}{AA}}}

\thanks{\textsuperscript{\faEnvelope[regular]}Corresponding Author: Amir Aly, University of Plymouth.}
}

\maketitle

\begin{abstract}
Early diagnosis and intervention for Autism Spectrum Disorder (ASD) has been shown to significantly improve the quality of life of autistic individuals. However, current diagnostic practices depend on behavioral assessments, which are prone to bias and fail to capture early manifestations of the disorder, making early diagnosis challenging. There is a need for objective biomarkers of ASD which can help improve early diagnostic accuracy. Extensive research has been conducted on creating machine learning models that classify ASD using resting-state functional Magnetic Resonance Imaging (fMRI) data. However, existing models lack interpretability. This research aims to improve the accuracy and interpretability of ASD diagnosis by creating a deep learning (DL) model that can not only accurately classify ASD but also provide explainable insights into its working. The dataset used is a preprocessed version of the Autism Brain Imaging Data Exchange (ABIDE). Our findings show a model that achieved a classification accuracy of 98.2\% and an F1 score of 0.97, while also highlighting critical brain regions differing between ASD and typical controls, with implications for early diagnosis and understanding of the neural basis of ASD. Additionally, a systematic evaluation of interpretability methods using the Remove And Retrain (ROAR) technique highlights the reliability of gradient-based approaches for fMRI data. These findings are validated by studies in the literature that use different datasets and modalities, confirming that the model actually learned characteristics of ASD and not just dataset artifacts.
\end{abstract}


\section{\textbf{Introduction}}
\label{sec:introduction}
\IEEEPARstart{A}{utism} Spectrum Disorder (ASD) is a complex neurodevelopmental condition characterised by impairments in social communication, interaction, and restricted, repetitive behaviors \cite{hirota_2023_autism}. Its presentation varies widely across affected individuals, impacting on quality of life. In the last 20 years, there has been a 787\% surge in ASD diagnosis increasing its prevalence from approximately 1\% to around 3\% of the population. This surge in diagnosis is thought to be due to the delay in recognition of the clinical symptoms and the lack of personnel to diagnose \cite{ayres_2017_a,saito_2020_prevalence,brentani_2013_autism}. Early diagnosis, ideally between the ages of two and five, can lead to a significantly better Quality of Life (QoL) for individuals with ASD \cite{estes_2015_longterm}. Studies also show that early intervention can improve IQ scores by up to 18 points and enhance language skills \cite{okoye_2023_early}, emphasising the critical need for early diagnosis and intervention. However, the average age of diagnosis remains around five years old, highlighting a crucial limitation in current diagnostic practices \cite{maenner_2023_prevalence}.

Specifically, current clinical ASD diagnostic practices lack effective early diagnosis methods. The method of diagnosing ASD in children uses behavioral analysis combined with clinical history, which is not effective for early diagnosis as it is based on identifying abnormal communication and social behaviors that may not emerge until the condition is well established \cite{mccarty_2020_early}. It is also subject to clinicians' biases and has historically led to under-diagnosing ASD particularly, in certain groups such as females \cite{estrin_2021_barriers}. To enhance our understanding of ASD and improve diagnostic accuracy, there is a need for objective, biologically-based markers of the condition.

The resting-state functional Magnetic Resonance Imaging (rs-fMRI) data of people with ASD has been analysed to show abnormalities in functional connectivity\footnote{Functional connectivity examines the correlation of activity between different brain regions. If two areas consistently show increases and decreases in activity together, they are considered functionally connected.} \cite{bi_2018_abnormal}. These \textit{differences} in \textit{distinctive} patterns in the functional connectivity of autistic individuals can serve as \textbf{\textit{biomarkers}} of ASD. However, recent research highlights the “idiosyncratic brain” concept, where functional connectivity patterns vary widely among individuals with ASD. This variability complicates the search for a universal biomarker and underscores the need for a tool that can analyse these subtle differences. (For additional detail on these neural connectivity patterns, see Section II-A.)

These differences can be learned using deep learning (DL) and be used to distinguish ASD and Typical Control (TC). Many recent studies have emerged using functional connectivity for ASD classification using machine learning, with some showing a high accuracy ($>$ 90\%) \cite{khodatars_2021_deep,wang_2019_identification}. While these models cannot replace clinical professionals, they can offer valuable decision support, but most current approaches lack in explainability. These limitations reduce trust and hinder clinical acceptance of the model in real-world scenarios \cite{shortliffe_2018_clinical}. To address this challenge, there is a large body of research focusing on approaches that enhance the interpretability of DL models in the medical domain \cite{teng_2022_a}. However, identifying which approach should be used with functional connectivity data or how they should specifically be used to identify biomarkers in ASD has not yet been explored.

This study explores the utilisation of DL methods to identify potential biomarkers for ASD through the analysis of resting-state fMRI data, aiming to develop and explain a predictive model that is both accurate and interpretable. We also benchmark a variety of interpretability methods using the Remove And Retrain (ROAR) technique \cite{hooker_2019_a} to identify the most effective method for interpreting DL models that are trained on functional connectivity fMRI data. 

This research has the potential to significantly improve our understanding of the neurobiological basis of ASD. This study also contributes to the growing field of explainable AI in healthcare, addressing the need for transparency and interpretability in medical AI systems, as required by regulatory bodies such as the FDA \cite{centerfordevicesandradiologicalhealth_2019_artificial}. Our findings are critically evaluated in the context of existing research and potential limitations are discussed. A comprehensive analysis of our model's performance and its implications for ASD research and diagnosis is provided. The regions highlighted by our model are compared with existing literature to verify whether our model genuinely learned about ASD characteristics or simply learned patterns specific to our dataset.

Successful application of the outcomes of this research could lead to improved access to early interventions, reduced costs of ASD diagnosis, and the development of more effective diagnostic tools. By enhancing our understanding of the neurobiological markers of ASD, our work aims to contribute to the development of more targeted interventions and improved quality of life for individuals with ASD and their families.

\section{\textbf{Related Work}}

\subsection{\textbf{Neurological Basis of ASD}}
Autism Spectrum Disorder (ASD) is associated with distinct yet heterogeneous patterns of functional connectivity in the brain, often manifesting as local over-connectivity and long-range under-connectivity \cite{belmonte_2004_autism, kana_2009_kana, kana_2014_brain}. Key regions implicated include the frontal cortex (executive functions, social cognition) \cite{just_2004_cortical}, temporal cortex (language processing) \cite{kana_2006_sentence}, amygdala (emotion processing) \cite{baroncohen_2000_the}, and the cerebellum (motor control, social cognition) \cite{wang_2014_the}. Abnormalities in large-scale networks, such as the Default Mode Network (DMN) and Salience Network (SN), are also linked to ASD \cite{kennedy_2008_functional, uddin_2013_salience}. However, considerable variability arises from the “idiosyncratic brain” concept \cite{hahamy_2015_the}, wherein each individual with ASD may exhibit unique functional connectivity patterns, making the search for a universal biomarker challenging. Moreover, symptom severity further modulates which regions display the greatest atypicality \cite{donovan_2016_the}. These factors collectively underscore the complexity of ASD’s neurobiological underpinnings and highlight the need for sophisticated methods—particularly deep learning (DL)—that can detect subtle and individualised patterns.

\subsection{\textbf{Deep Learning Approaches for ASD}}
Deep learning has emerged as a promising tool in ASD research due to its ability to extract complex, nonlinear features from high-dimensional data such as fMRI \cite{lecun_2015_deep, geenjaar_2022_deep, xu_2023_deep}. Various architectures (e.g., CNNs, RNNs, GNNs, and autoencoders) have been explored to improve diagnostic accuracy and potentially enable early detection \cite{geenjaar_2022_deep,xu_2023_deep,abessadok_2022_graph}. However, existing studies vary widely in sample size, accuracy, and interpretation strategies.

For example, a study \cite{li_2018_brain} applied a deep neural network classifier to a small dataset of fewer than 130 samples, achieving accuracies between 85.3\% and 87.1\%. Although they introduced a single interpretability method, they did not determine whether the highlighted features were consistent with known ASD-related neurobiology. In a separate study, \cite{wang_2019_identification} leveraged over 1,000 samples from the Autism Brain Imaging Data Exchange (ABIDE) to attain a high accuracy of 93.59\% using a Stacked Sparse Autoencoder (SSAE), but offered no mechanism to clarify which specific brain connections contributed most to classification. Independently, another group of authors \cite{wang_2021_autistic} focused on biomarker discovery and employed an interpretability method, yet their accuracy was lower at around 72\% and they did not cross-reference the discovered biomarkers with established ASD findings. Collectively, these studies demonstrate the potential of DL in ASD classification, particularly for large-scale data, but also reveal critical shortcomings: either interpretability is absent, or when present, accuracy is low and it is rarely corroborated with well-established ASD findings from the neuroscience literature.

\subsection{\textbf{Feature Analysis and Interpretability in ASD Research}}
Interpretability methods bridge the gap between raw classification accuracy and meaningful clinical insights. Gradient-based algorithms, such as Layer-wise Relevance Propagation (LRP) \cite{montavon_2018_methods} and DeepLift \cite{shrikumar_2017_learning}, and perturbation-based tools like SHAP \cite{lundberg_2017_a} or LIME \cite{ribeiro_2016_}, have been used successfully in other medical contexts \cite{bhle_2019_layerwise,sundararajan_2017_axiomatic,mageshpavanrajkumar_2020_an}. Yet, ASD research presents unique challenges: the heterogeneous presentation of ASD may mean that each subgroup has distinct connectivity markers, there is no absolute ground truth for “crucial” regions in ASD, and most existing studies that do employ interpretability methods fail to confirm whether their highlighted regions align with validated ASD neurobiology \cite{salahuddin_2022_transparency, riaz_2020_deepfmri}. Further compounding this is the lack of a standard framework for benchmarking interpretability in ASD-specific fMRI data, leaving researchers to rely on a single method without broader validation or comparisons.

\subsection{\textbf{Gaps in Current Research}}
While deep learning models can offer high accuracy in ASD classification, several persistent gaps limit both clinical utility and scientific understanding:
\begin{enumerate}
    \item \textbf{Lack of Transparent DL Models:} Many high-accuracy studies \cite{wang_2019_identification} do not explain which brain regions or connections drive their decisions, leading to “black box” models.
    \item \textbf{Limited Interpretability Benchmarking:} The few studies that do focus on interpretability only use a single approach to highlight features, leaving uncertainty about whether alternate methods would reveal or confirm the same features.
    \item \textbf{Sparse Validation Against Neuroscience:} Even when interpretability is attempted, few works cross-reference identified biomarkers with established ASD findings, raising doubts about clinical relevance \cite{li_2018_brain, wang_2021_autistic}.
\end{enumerate}

These challenges highlight the need for an approach that combines high diagnostic performance with robust interpretability and neuroscientific validation. In this paper, we address these gaps by:
\begin{itemize}
    \item Developing a DL pipeline that can classify and interpret ASD with high accuracy.
    \item Evaluating multiple interpretability methods in a systematic benchmark.
    \item Cross-referencing the identified brain regions with established ASD research to ensure clinical and biological relevance.
\end{itemize}

By uniting rigorous performance evaluation, interpretability, and domain validation, we address the critical need for a transparent system that can unveil meaningful biomarkers and build clinical trust.

\section{\textbf{Methodology}}

\subsection{\textbf{Data Collection and Preprocessing}}

\subsubsection{\textbf{Dataset}}

This study used the Autism Brain Imaging Data Exchange (ABIDE) dataset \cite{martino_2014_the}, which comprises resting-state fMRI data from 539 autistic and 573 typically developing individuals aged 7--64 years from 16 international imaging sites. Note that the sample's gender imbalance (85.2\% male) may limit generalisability \cite{wachinger_2018_detect}. The data is anonymised in accordance with standard ethical protocols and were made available under the 1000 Functional Connectomes Project (FCP) and International Neuroimaging Data-Sharing Initiative (INDI).

\subsubsection{\textbf{Preprocessing}}

The Preprocessed Connectomes Project (PCP) openly shares ABIDE data preprocessed via four pipelines---the Connectome Computation System (CCS) \cite{xu_2015_a}, the Configurable Pipeline for the Analysis of Connectomes (CPAC) \cite{craddock_2013_towards}, the Data Processing Assistant for Resting-State fMRI (DPARSF) \cite{yan_2010_dparsf}, and the NeuroImaging Analysis Kit (NIAK) \cite{simexplab_2016_niak}. These pipelines are intended to mitigate center-specific artifacts. Because no single pipeline is universally optimal, we trained our model on all four to compare performance \cite{_2012_abide}. Additionally, analysing data from multiple pipelines helps us identify the core regions that remain consistent regardless of preprocessing choices, ensuring that any discovered biomarkers are pipeline-independent. Full details on these pipelines are available in the PCP documentation \cite{_2012_abide}.

Each pipeline applies standard low-level preprocessing steps, such as brain extraction, realignment (motion correction), and slice timing correction \cite{craddock_2013_towards}. Further, there has been significant debate in the literature about whether Framewise Displacement (FD)\footnote{Framewise Displacement (FD) is a measure of how much head movement occurs during the MRI scan} filtering influences ASD classification performance. Some influential studies state that an FD value \textgreater 0.2\,mm corrupts the fMRI data \cite{power_2014_methods,power_2012_spurious}, leading to decreased accuracy, whereas another study \cite{iidaka_2015_resting} contends that motion does not appreciably affect performance. We conducted experiments both with and without FD filtering to clarify this methodological issue.

Applying an FD threshold of 0.2\,mm yielded 884 usable samples (408 ASD, 476 TC). These were drawn from multiple international sites and scanners, which increases data diversity but may introduce additional variability.

\subsubsection{\textbf{Parcellation}}

Brain images were then parcellated with the Automated Anatomical Labelling (AAL) atlas, which divides the brain into 116 Regions of Interest (ROIs). We selected the AAL atlas for its balance of anatomical detail and manageable complexity, as recommended in similar ASD studies \cite{wang_2019_identification, liu_2024_madeforasd}. This produced an $N\times t$ data array (where $N=116$ ROIs, $t$ is the temporal dimension) for each of the 884 subjects, providing the foundation for subsequent feature extraction.

\subsection{\textbf{Feature Extraction}}

To reveal the patterns of coordination between brain regions, each fMRI time-series was converted into a functional connectivity matrix (FCM) by calculating Pearson’s correlation coefficient $R(i,j)$ between the intensity time series from every pair of ROIs. We then applied Fisher’s Z transformation to normalise these correlation coefficients. Finally, to reduce redundancy, we flattened the lower triangular of each symmetric FCM $(116\times116)$, yielding $6670$ features per sample.

Creating a feature vector from the FCM has been thoroughly researched and used in DL models, however, when using them directly for autism diagnosis, the DL models peaked at an accuracy of 66.8\% \cite{almuqhim_2021_asdsaenet,li_2018_a}. This implies that these DL models suffer from the high-dimensionality and low number of samples problem. To efficiently use DL models, feature dimensionality must be reduced, or sample size must be increased. However, increasing sample size is infeasible due to reasons including high MRI costs and difficulty in participant recruitment. 

To address this, we employed Support Vector Machine with Recursive Feature Elimination (SVM-RFE) for dimensionality reduction. We chose SVM-RFE (over unsupervised methods like PCA) for its task-specific feature selection and proven performance in similar studies  \cite{wang_2019_identification}, ultimately reducing the feature set from 6,670 to 1,000 features per sample. This mitigates the computational burden and helps the model focus on the most discriminative functional connections.

\begin{figure*}
    \centering
    \includegraphics[width=\linewidth]{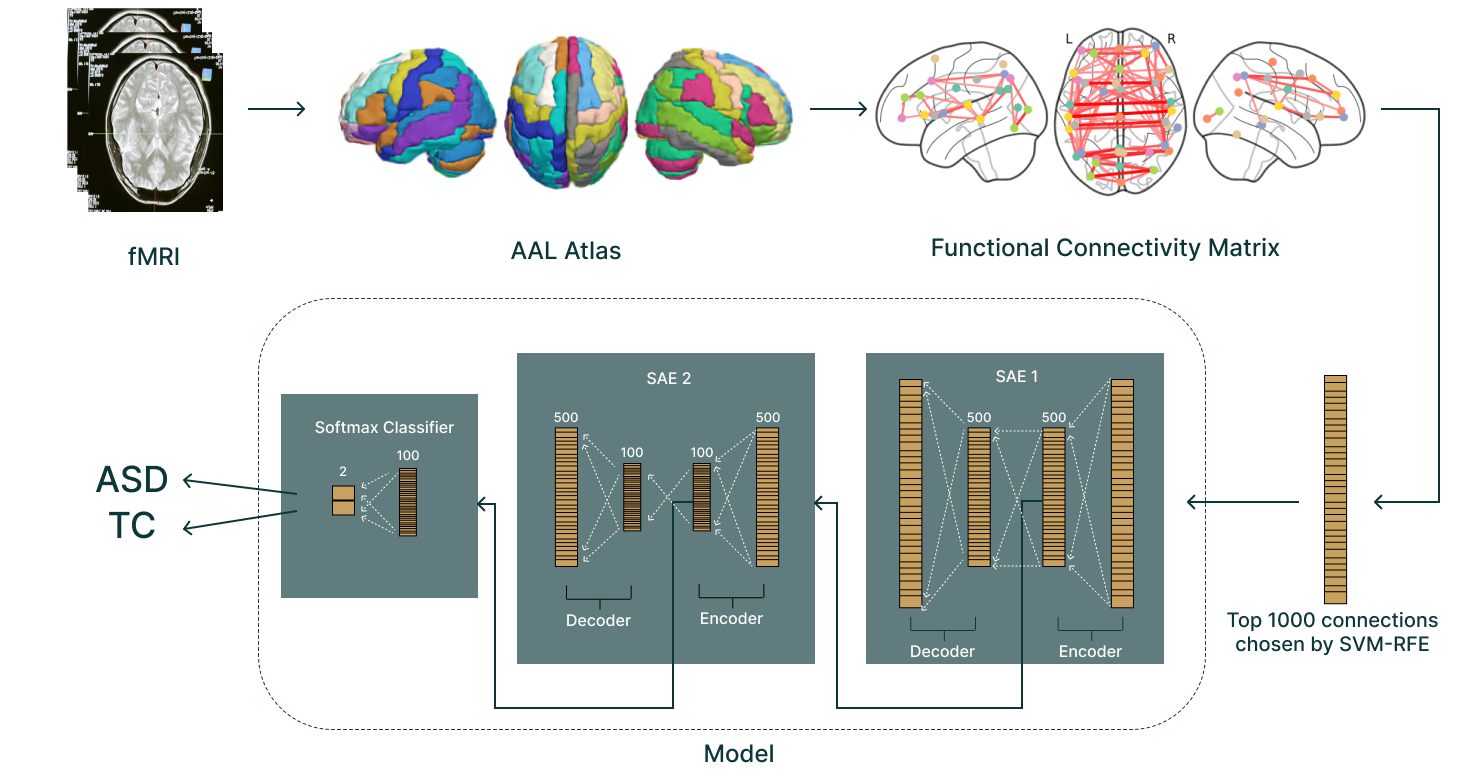}
    \caption{The overall architecture of the model, incorporating both the SSAE and softmax classifier.}
    \label{fig:model-arch}
\end{figure*}

\subsection{\textbf{Deep Learning Model}}

This study utilises a DL architecture (Figure \ref{fig:model-arch}) comprising a Stacked Sparse Autoencoder (SSAE) combined with a softmax classifier to classify ASD using functional connectivity data. Autoencoders learn a compressed representation of the input, and the \textit{sparse} constraint encourages focus on only the most distinctive biomarkers, which is important for high-dimensional functional connectivity data.

Our choice of a two-layer Stacked Sparse Autoencoder (SSAE) with ReLU activation and KL divergence\footnote{Kullback-Leibler (KL) divergence is a measure from information theory that quantifies how different one probability distribution is from another} sparsity penalty is motivated by both practical and empirical considerations \cite{li_2018_a,wang_2019_identification}. We found that two hidden layers strike an optimal balance between expressive power and training complexity, especially given our sample size. Hyperparameter tuning indicated that deeper networks did not improve performance, so this design effectively balances model expressiveness with the risk of overfitting. ReLU activations have shown robust performance in similar high-dimensional neuroimaging tasks, avoiding saturation issues observed in sigmoid/tanh functions. Crucially, the KL divergence term enforces neuron-level sparsity, highlighting discriminative features while mitigating overfitting \cite{almuqhim_2021_asdsaenet}, as captured by the following loss function $L(x,\hat{x})$ between input $x$ and its reconstruction $\hat{x}$:

\begin{equation}
L(x, \hat{x}) = MSE + \beta \times KL(p \| \hat{\rho_j})
\label{eq:loss_fn}
\end{equation}

where $MSE$ is the mean squared error between input $x$ and its reconstruction $\hat{x}$, $\rho$ is the desired sparsity, $\hat{\rho}_j$ is the average activation of the $j$-th hidden neuron, and $\beta$ controls the weight of the sparsity penalty.

Our SSAE consisted of two autoencoder layers that used the ReLU activation function. The first reduces 1000 input features to 500, and the second further compresses to 100 neurons. The output was then fed into a softmax classifier that was trained using cross-entropy loss for ASD/TC classification. This is visualised in Figure (\ref{fig:model-arch}).

\subsection{\textbf{Training and Evaluation}}

The model was trained in two stages—unsupervised pre-training followed by supervised fine-tuning—to combine the strengths of both approaches. We used stratified 5-fold cross-validation to balance computational cost with variance reduction, given our dataset of 884 samples. Within each fold, the data was split into 64\% training, 16\% validation, and 20\% testing, ensuring distinct subsets at every stage. A public repository with our code will be provided to facilitate reproducibility.

\subsubsection{\textbf{Unsupervised Pre-training}}

The SSAE underwent unsupervised pre-training using the greedy layer-wise training approach \cite{bengio_2006_greedy}. This sequential approach allows each autoencoder layer to learn increasingly abstract representations of the input data. SAE 1 was trained first and its compressed hidden layer representation then served as input for training SAE 2. Each layer used Mean Squared Error (MSE) combined with a sparsity penalty via KL Divergence for reconstruction fidelity \cite{li_2018_a,yin_2022_a,wang_2019_identification}. We employed the Adam optimiser \cite{kingma_2014_adam} throughout, chosen for its adaptive learning rate and rapid convergence.

\subsubsection{\textbf{Supervised Fine-Tuning}}

Once pretraining was complete, the encoder layers were combined with a softmax classifier. In contrast to approaches that fine-tune only the classifier, we fine-tuned both the encoder and classifier parameters together to enhance discrimination between ASD and TC \cite{almuqhim_2021_asdsaenet,li_2018_a,yin_2022_a}. This supervised phase again used Adam, with hyperparameters\footnote{Hyperparameters: Pre-training - 50 epochs, learning rate of 0.001, weight decay of 0.0001, batch size of 128, KL Divergence parameters: $\rho$ = 0.2, $\beta$ = 2. Fine-tuning - 50 epochs, learning rate of 0.0001, weight decay of 0.0001, batch size of 128.} adjusted to avoid overfitting while refining the learned representations for classification.

\subsubsection{\textbf{Performance Evaluation}}

We report accuracy, sensitivity, specificity, precision, and F1-scores, each averaged across the five k-folds. This ensures a thorough assessment of model performance beyond simple accuracy. Additionally, we monitored training and validation losses to detect overfitting, providing insight into whether the model surmounts typical performance limitations reported in prior ASD classification work.

\subsection{\textbf{Feature Analysis and Interpretation}}

This section explores the different methods that were used to extract meaningful features used by the model for its diagnosis. Understanding the basis of its decisions is essential for trust and clinical translation.

\subsubsection{\textbf{Interpretability Methods}}

This section explores the interpretability methods employed to reveal how the model arrives at its decisions. Interpretable features are crucial for clinical acceptance, as they help clarify the neural mechanisms underlying ASD classification and thus support potential clinical translation.

We evaluated seven approaches (gradient- and perturbation-based) to capture different perspectives of feature importance. Specifically, we used Integrated Gradients \cite{sundararajan_2017_axiomatic}, LIME \cite{ribeiro_2016_}, SHAP \cite{lundberg_2017_a}, DeepLift \cite{shrikumar_2017_learning}, DeepLiftShap \cite{chen_2021_explaining}, GradientShap \cite{lundberg_2017_a}, and GuidedBackprop \cite{springenbergjosttobias_2014_striving}. Gradient-based methods (Integrated Gradients, DeepLift, DeepLiftShap, GradientShap) directly quantify how changes in input features affect predictions. LIME employs local perturbations to approximate model decisions, whereas SHAP uses a game-theoretic framework to attribute feature importance. GuidedBackprop highlights salient input features via positive gradients. Because each technique might emphasise different aspects of the data, benchmarking them was essential for trustworthy interpretation.

\subsubsection{\textbf{Remove and Retrain (ROAR)}}

We systematically compared these methods using the Remove and Retrain (ROAR) framework \cite{hooker_2019_a}. First, each interpretability technique ranks the input features by importance. Next, we replace the most influential features (above a chosen threshold) with an uninformative value (zero)\footnote{Zero was chosen as the uninformative value because the input features are correlation coefficients and zero correlation indicates that no positive or negative relationship exists between the two variables. Hence, zero is uninformative in the context of functional connectivity.}. Finally, we retrain the model on this modified dataset and measure the drop in classification performance. Larger performance decreases indicate that the removed features were indeed critical for the model’s predictive capability.

\subsubsection{\textbf{Important Regions of Interest (ROIs)}}

\begin{figure*}
    \centering
    \includegraphics[width=1\linewidth]{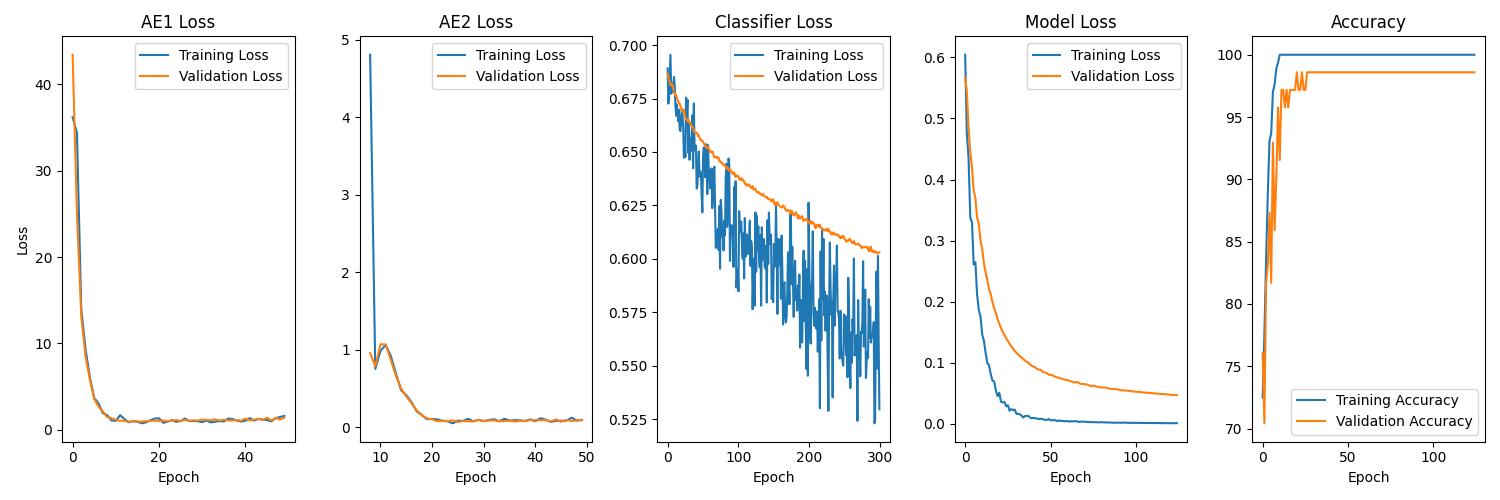}
    \caption{Loss graphs during first K-Fold. AE1, AE2, and classifier loss represent reconstruction loss during unsupervised pre-training. Model loss and accuracy refer to the supervised fine-tuning stage.}
    \label{fig:loss-graph}
\end{figure*}

\begin{figure}
    \centering
    \includegraphics[width=0.65\linewidth]{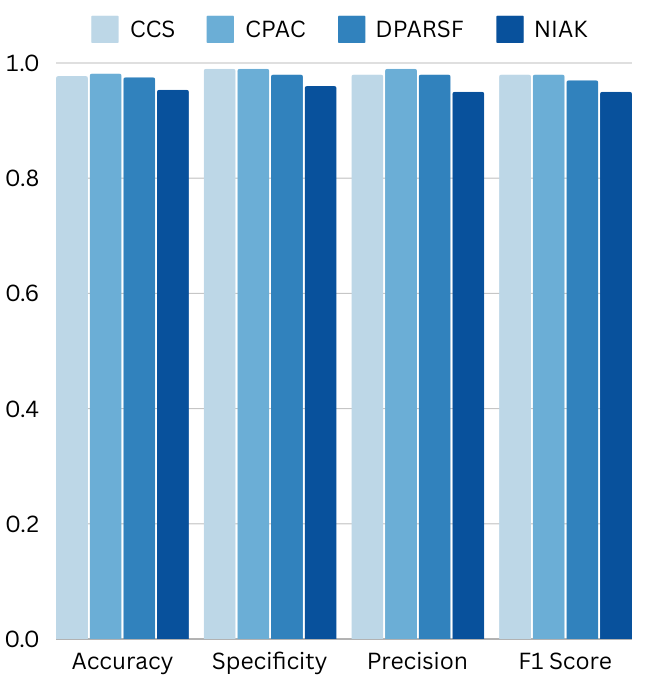}
    \caption{Performance of the model across different preprocessing pipelines averaged across 5 K-folds.}
    \label{fig:k_fold_performance}
\end{figure}

\begin{figure*}
    \centering
    \includegraphics[width=0.9\linewidth]{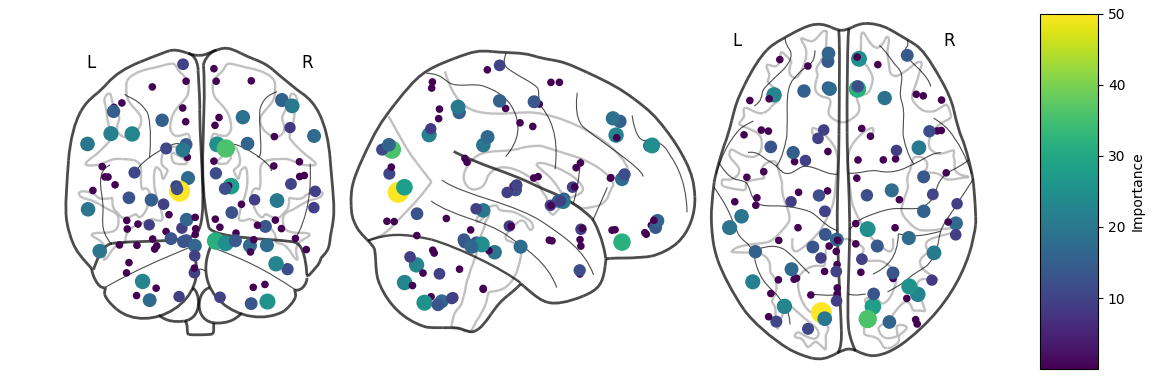}
    \caption{Important ROIs (yellow) identified by the Integrated Gradients approach when using DPARSF pipeline.}
    \label{fig:roi_dparsf}
\end{figure*}

After determining which features (correlation coefficients between brain regions) were most important, we aggregated them at the region level. Specifically, we identified how often each Region of Interest (ROI) appeared among top-weighted features and then mapped these important ROIs to their anatomical labels, visualising the results (Figure \ref{fig:roi_dparsf}). We then mapped the features across the different pipelines to Brodmann Areas (Table \ref{tab:brain_regions}) and analysed the overlapping features (Figure \ref{fig:broadman_area}). This analysis enabled us to pinpoint the key brain areas driving ASD classification and to compare our findings against established ASD-related neurobiology.

\section{\textbf{Results and Discussion}}

The model achieved state-of-the-art performance on the ABIDE dataset, with a maximum accuracy of 98.2\% across three different preprocessing pipelines, exceeding the previous benchmark of 93.59\% (Table \ref{tab:compared-to-other-papers}). Our overall metrics were equally strong (specificity: 0.98, precision: 0.98, F1-score: 0.97). Below, we detail performance and interpretability results, emphasising the clinical significance of the identified regions.

\begin{table}[]
\centering
\resizebox{0.35\textwidth}{!}{%
\begin{tabular}{|c|c|}

\hline
\textbf{Model}         & \textbf{Accuracy(\%)} \\ \hline
Pavithra et al. (2023) \cite{pavithra_2023_identification} & 85.0                  \\ \hline
Bhandage et al. (2023) \cite{bhandage_2023_autism} & 92.4                  \\ \hline
Wadhere et al. (2023)  \cite{wadhera_2023_a} & 88.1                  \\ \hline
Wang et al. (2019)     \cite{wang_2019_identification} & 93.5                  \\ \hline
\textbf{Our model}             & \textbf{98.2}         \\ \hline

\end{tabular}%
}
\caption{Accuracy comparison with studies that utilise fMRI modality and have datasets comprising more than 500 samples.} 
\centering
\label{tab:compared-to-other-papers}
\end{table}

\begin{table}
    \centering
    \begin{tabular}{|c|c|}
        \hline
        \textbf{Model} & \textbf{Accuracy} \\ \hline
        Model without FD Filtering & 91\% \\ \hline
        Model with FD Filtering & 98.2\% \\ \hline
    \end{tabular}
    \caption{Model accuracy with and without framewise displacement filtering of samples having $>$ 0.2mm FD.}
    \label{tab:model_acc_tab1}
\end{table}

\subsection{\textbf{Model performance}}

Filtering scans with framewise displacement (FD) \textgreater 0.2,mm boosted accuracy from 91\% to 98.2\% (Table \ref{tab:model_acc_tab1}), underscoring the importance of motion correction \cite{power_2012_spurious,power_2014_methods}. This 7\% jump aligns with the autoencoders’ struggles to reconstruct high-FD samples, although some studies report minimal effects of FD \cite{iidaka_2015_resting}. Motion correction is thus essential to ensure consistent biomarker detection, an especially relevant point for younger populations where motion artifacts may be more frequent and yet early diagnosis is most critical \cite{mccarty_2020_early}. 

Among the four preprocessing pipelines (CCS, CPAC, DPARSF, and NIAK), DPARSF achieved the highest accuracy (98.2\%), while CCS and CPAC trailed within 1\% and NIAK was about 4\% lower (Figure \ref{fig:k_fold_performance}). In addition, ABIDE’s multi-center nature introduces site-specific noise from different scanners and protocols. By evaluating four distinct preprocessing pipelines, we partially addressed these variations; however, the relatively lower performance of the NIAK pipeline suggests residual sensitivity to cross-site heterogeneity for NIAK. Future investigations could explore advanced domain adaptation or harmonisation strategies to further mitigate site-based variability.

To check for overfitting, we monitored training and validation loss curves (Figure~\ref{fig:loss-graph}), observing stable convergence without divergence. Moreover, the important ROIs identified by our model closely match independent neuroimaging findings \cite{gandal_2022_broad,robertson_2014_global}, suggesting that it captures genuine ASD-relevant patterns rather than memorising dataset artifacts. Nonetheless, testing on external data is a near-future objective to confirm out-of-distribution generalisation.
Observing Figure (\ref{fig:loss-graph}), the autoencoders (AE1 and AE2 loss) consistently reconstruct the data during unsupervised pre-training, suggesting that broad, shared features are captured prior to differentiating ASD from TC. In the subsequent supervised fine-tuning phase, cross-entropy loss (Model loss) drops steadily for both training and validation sets, indicating the network effectively learns the subtle differences between ASD and TC without overfitting. This hybrid approach—unsupervised pre-training plus supervised fine-tuning—appears to capture robust, clinically relevant patterns in functional connectivity, consistent with findings on multi-stage training in neuroimaging tasks \cite{albadawyehaba_2018_deep}.

\subsection{\textbf{Feature Importance}}

\subsubsection{\textbf{ROAR Benchmark}}

\begin{figure}
    \centering
    \includegraphics[width=1\linewidth]{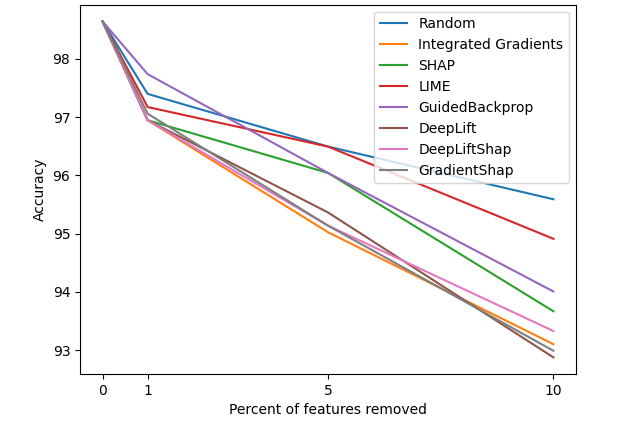}
    \caption{ROAR Analysis, shows how removal of top 1\% of features leads to steeper accuracy drop for the Integrated Gradients, DeepLift, DeepLiftShap and GradientShap methods.}
    \label{fig:roar}
\end{figure}

The ROAR analysis identified which interpretability methods are reliable. It showed that Integrated Gradients, Deep
fLift, DeepLiftShap, and GradientShap methods all had similar results and showed the steeper drop in accuracy when the initial features were removed, indicating that the gradient-based methods identified the most discriminative features required for ASD classification (Figure \ref{fig:roar}). The Integrated Gradients approach showed the steepest drop in accuracy and was therefore assumed the most reliable method. Notably, the LIME approach did not perform much better than the random baseline, indicating that feature perturbation-based techniques do not perform well in the functional connectivity field. This contrasts with results found in the original ROAR study \cite{hooker_2019_a} that showed that the Integrated Gradients approach performed similarly to the random baseline, further emphasising that the best-performing interpretability methods change depending on the model architecture and the data's modality.

\subsubsection{\textbf{Feature Analysis}}
\label{sec:feature-analysis}
Feature analysis using the Integrated Gradients approach revealed a set of consistent brain connectivity patterns highly discriminative for the classification of ASD across the different preprocessing pipelines, one of which can be observed in Figure (\ref{fig:roi_dparsf}). 

Important brain areas (parcellated per the AAL atlas), identified by the interpretation method are shown in columns CCS, CPAC, and DPARSF in Table (\ref{tab:brain_regions}). The regions are arranged in descending order of importance score, with their corresponding Brodmann Areas \cite{garey1999brodmann}. The most consistently important regions across pipelines are areas 17 (Primary visual cortex) and 18 (Secondary visual cortex), with areas 41 and 42 (Auditory cortex) also showing significance. 

The consensus results across the three pipelines suggest that ROIs involved in visual processing are central to the classification of participants as autistic versus controls. The occipital lobe identifies three visual regions of particular importance: the calcarine sulcus, cuneus, and superior occipital gyrus. 

Notably, all three pipelines rank the calcarine sulcus as the most important area. This finding is particularly significant as it aligns with current evidence from multiple research dimensions, including genetic studies that have independently highlighted the importance of Brodmann areas 17 and 18 in autism \cite{gandal_2022_broad}. Other research dimensions including studies on global motion perception deficits \cite{robertson_2014_global} and atypical gamma oscillations \cite{orekhova_2023_gamma} show ASD being associated to the primary visual cortex. Our model's identification of this region corroborates growing evidence that fundamental differences in early visual processing may play a crucial role in ASD, potentially influencing broader social and behavioral symptoms.

Similarly, the cuneus (part of Brodmann area 17) emerges as a significant region. Recent research has shown that autistic individuals have reduced connectivity between the brainstem and the cuneus compared to their non-autistic co-twins, potentially indicating changes in the low-level visual pathway that may affect social information processing and low-level perception \cite{cheng_2015_autism}. Furthermore, fMRI and eye-tracking studies have implicated the cuneus in social dysfunction in autism \cite{xiao_2023_atypical}. 

The pipelines also agree on the importance of the angular gyrus, located in the posterior part of the inferior parietal lobe, which is involved in language processing. This finding aligns with the known language and communication difficulties often observed in ASD \cite{monk_2009_abnormalities}. 

\begin{figure}
    \centering
    \includegraphics[width=1\linewidth]{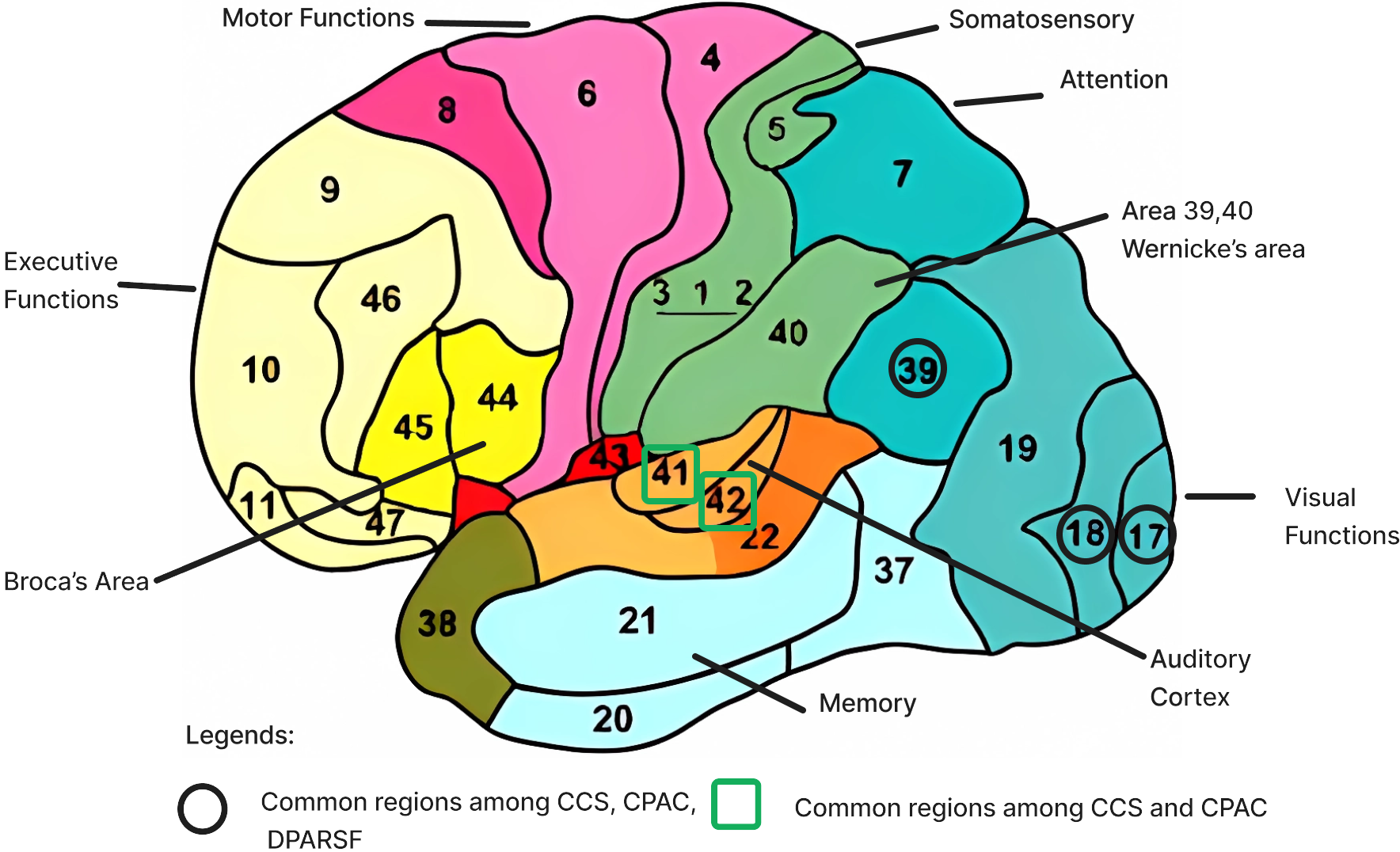}
    \caption{Shows the Brodmann's area\cite{broadmann_area_fig}. The different colors represent different brain functions. The common important brain regions from the Table (\ref{tab:brain_regions}) are highlighted. }
    \label{fig:broadman_area}
\end{figure}

The convergence of our machine learning-based findings with these different studies provides strong validation that our model has indeed captured genuine neurobiological markers of ASD, rather than dataset-specific patterns. These results also demonstrate its potential to corroborate and extend existing knowledge in autism research. The consistency of our findings across different preprocessing pipelines and their alignment with diverse research approaches (genetic, neuroanatomical, and functional) highlights the robustness of our method and its potential for advancing our understanding of the neurobiological basis of ASD. Figure (\ref{fig:broadman_area}) highlights these important regions, providing a visual representation of the key areas identified by our analysis.

Importantly, while these top ROIs offer potential insight, it is crucial to remember that ASD is a complex disorder. Significant patterns likely arise from a combination of interacting brain regions. Further investigation might focus on broader network-level differences rather than isolated ROIs. It's noteworthy that our dataset included a variety of ASD severity cases, and the identified biomarkers were common among all of them. This suggests that these visual processing regions may represent a fundamental neurobiological signature of ASD, potentially present across the spectrum of severity.

\begin{table*}%[H]
\resizebox{\textwidth}{!}{%
\begin{tabular}{|c|c|c|c|c|c|c|}
\hline
\textbf{CCS}               & \textbf{Brodmann Area} &  \textbf{CPAC}                 & \textbf{Brodmann Area} &  \textbf{DPARSF}                  & \textbf{Brodmann Area} \\ \hline
Calcarine\_L      & 17            &  Calcarine\_R         & 17            &  Calcarine\_L            & 17            \\
Occipital\_Sup\_R & 17            &  Cuneus\_R            & 17,18        &  Rectus\_R               & 11,12         \\
Occipital\_Mid\_R & 17            &  Occipital\_Sup\_R    & 17           &  Cuneus\_R               & 17,18         \\
Cuneus\_R         & 17,18        &  Frontal\_Inf\_Tri\_R & 44,45       &  Temporal\_Mid\_L        & 21            \\
Angular\_R        & 39            &  Angular\_L           & 39           &  Cerebelum\_Crus1\_R     &       -        \\
Cerebelum\_8\_L   &       -        &  Cerebelum\_8\_R      &      -        &  Calcarine\_R            & 17            \\
Thalamus\_R       & 41,42         &  Cerebelum\_Crus2\_L  &        -      &  Cerebelum\_3\_R         &           -    \\
Calcarine\_R      & 17            &  Calcarine\_L         & 17           &  Temporal\_Inf\_L        & 20            \\
Insula\_R         & 13,16         &  Lingual\_L           & 17           &  Occipital\_Sup\_R       & 17            \\
SupraMarginal\_R  & 40            &  ParaHippocampal\_L   & 34           &  Frontal\_Sup\_Medial\_R & 11,12         \\
Parietal\_Inf\_L  & 5             &  Precentral\_R        & 4            &  Frontal\_Sup\_L         & 11,12         \\
Heschl\_R         & 41            &  Thalamus\_L          & 41,42        &  Angular\_L              & 39            \\ \hline
\end{tabular}%
}
\caption{The top ROIs identified by each pipeline (CCS, CPAC and DPARSF) in decreasing order of importance alongside its corresponding Brodmann Area.}
\label{tab:brain_regions}
\end{table*}

\section{\textbf{Conclusion}}

This research significantly advanced the field of ASD classification using deep learning models applied to resting-state fMRI data, addressing several key gaps in current literature. Our model achieved exceptional performance with an average accuracy of 98.2\% using the DPARSF pipeline, surpassing the previous benchmark of 93.59\% \cite{wang_2019_identification}. This addresses the critical gap of lacking an explainable model with over 95\% accuracy in ASD classification. The model's success stems from its hybrid architecture, combining unsupervised pre-training and supervised fine-tuning, and the removal of corrupt data. This helped the model effectively capture complex functional connectivity patterns characteristic of ASD.

Unlike many previous studies that focused solely on accuracy, we provided comprehensive evaluation metrics including sensitivity, specificity, precision, and F1 scores. This thorough approach offers a more robust assessment of the model's performance and addresses the limitation of narrow evaluation practices in the field. Our study also revealed the significant effect of head movement on the precision of ASD diagnosis, contrasting with some previous findings and highlighting the importance of motion correction in fMRI studies of ASD.

We pioneered the use of ROAR (Remove And Retrain) to benchmark different deep learning interpretability methods in the context of functional connectivity data. This novel application revealed that gradient-based methods, particularly Integrated Gradients, are most suitable for this type of data, filling a significant gap in understanding the reliability of various interpretability approaches in ASD research.

Our feature analysis highlighted the importance of visual processing regions, particularly the visual processing regions (calcarine sulcus and cuneus), in ASD classification. These findings were consistent across different preprocessing pipelines and aligned with independent genetic studies \cite{gandal_2022_broad}, providing strong validation of our model's ability to capture genuine neurobiological markers of ASD rather than dataset-specific patterns. Notably, our identified biomarkers were common among a variety of ASD severity cases in our dataset, suggesting these visual processing regions may represent a fundamental neurobiological signature of ASD across the spectrum of severity.

In summary, this research has made significant strides in the application of deep learning to ASD classification and biomarker identification. By developing an accurate, explainable model and rigorously benchmarking interpretability methods, we have advanced the technical aspects of ASD diagnosis and contributed to the neurobiological understanding of the disorder. The consistent identification of visual processing regions as key discriminators across different severity levels of ASD opens new avenues for research and potentially for clinical applications. While challenges remain, this work represents a substantial step forward in the quest for objective, reliable, and interpretable methods for ASD diagnosis and characterisation.

\section{\textbf{Limitations and Future Directions}}

Despite the promising results of this study, several limitations must be acknowledged, each of which points to important directions for future research. A primary limitation is the binary nature of our classification model, which currently only distinguishes ASD from typical controls. By focusing on a binary classification, we risk overlooking the unique characteristics of other neurodevelopmental conditions that could share some of the same features.

Although we report strong results in the ABIDE dataset, testing on other multi-center, ethnically diverse rs-fMRI datasets or prospectively acquired clinical data is essential to confirm real-world portability. We plan to extend our pipeline to these data in future work, especially considering the considerable resource requirements for new fMRI scans. In principle, if new data uses a similar preprocessing pipeline and parcellation scheme, the model would not need retraining. Otherwise, we anticipate that partial or complete retraining will be required.

Another future direction would be to develop a model that can highlight important regions for ASD classification on a per-MRI-scan basis, rather than only analysing overall group differences. This individualised approach could enhance interpretability and identify brain regions that are relevant to a specific patient's presentation, potentially helping clinical trust and targeted intervention strategies.

Additionally, a multimodal approach holds significant promise. Integrating behavioural, cognitive, genetic, and structural imaging data alongside functional connectivity could provide a more holistic and nuanced picture of ASD. This approach may lead to the identification of ASD subtypes or personalised profiles, enabling more effective interventions. Finally, collaborative research is needed between neuroscientists, AI researchers, and clinicians to design clinically relevant studies and ensure that AI-driven diagnostic methods account for real-world complexities, including data diversity, and ethical deployment of AI in clinical settings.

\ifCLASSOPTIONcaptionsoff
  \newpage
\fi

\bibliographystyle{IEEEtran}
\bibliography{refs}

\end{document}